\documentclass[a4paper]{article}
\pdfoutput=1
\usepackage{arxiv}
\usepackage{subcaption}
\usepackage{calc}
\usepackage[utf8]{inputenc}
\usepackage[T1]{fontenc}   
\usepackage{url}         
\usepackage{booktabs}     
\usepackage{amsfonts}     
\usepackage{nicefrac}       
\usepackage{microtype}      
\usepackage{algorithm2e}
\usepackage{graphicx}

\title{Imperfect Oracles: The Effect of Strategic Information on Stock Markets}
\author{ \href{https://orcid.org/0000-0002-2185-7404}{\includegraphics[scale=0.06]{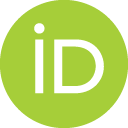}\hspace{1mm}Miklos Borsi}
	\thanks{https://orcid.org/0000-0002-2185-7404}\\
	Department of Computer Science\\
	University of Bristol\\
	Bristol, United Kingdom \\
	\texttt{bmiki97@gmail.com} \\
}

\begin{document}
\maketitle

\begin{abstract}
Modern financial market dynamics warrant detailed analysis due to their significant impact on the world. This, however, often proves intractable; massive numbers of agents, strategies and their change over time in reaction to each other leads to difficulties in both theoretical and simulational approaches.
Notable work has been done on strategy dominance in stock markets with respect to the ratios of agents with certain strategies. Perfect knowledge of the strategies employed could then put an individual agent at a consistent trading advantage. This research reports the effects of imperfect oracles on the system - dispensing noisy information about strategies - information which would normally be hidden from market participants. 
The effect and achievable profits of a singular trader with access to an oracle were tested exhaustively with previously unexplored factors such as changing order schedules. Additionally, the effect of noise on strategic information was traced through its effect on trader efficiency.
\end{abstract}


\section{Introduction}
\label{sec:introduction}

\noindent A line of research was started by Vernon Smith's experiments in a paper \cite{smith1962experimental} studying the trading behaviour of humans and allocative efficiency of the market as a whole in a Continuous Double Auction, the style of market mechanism used in almost all financial exchanges around the world. This work was groundbreaking for its experimental approach to economic theory which previously often held unclear or inaccurate prior beliefs about its claims. For example, the number of participating agents in a double auction that achieves a good equilibrium was defined as ``numerous'' or with the common mathematical approximation ``close to infinite''. In the real, reproducible situation however, the ``invisible hand'' of the market was shown to be in effect from as little as 8 agents, 4 buyers and 4 sellers, who were also untrained for the situation.

Initial work by Gode and Sunder \cite{godesunder} indicated that traders with practically zero intelligence - but some constraints - can produce the invisible hand effect as well. This was later proven by Cliff to be a mere byproduct of the underlying probability distribution and the supply-demand curves used \cite{cliff1997zero}. 

Cliff proposed an algorithm called Zero Intelligence Plus that emulated the previous attempt to make a minimally intelligent trading algorithm. In IBM's 2001 experiments \cite{ibm2001} algorithmic traders were able to outperform humans and made 7\% more profit on average. The algorithms included ZIP, Kaplan's Sniper and Gjerstad-Dickhaut \cite{gd1998} \cite{gdx2002} \cite{kaplansniperintro} \cite{kaplantournament}. 

The next addition to the algorithmic roster came in 2008 with Vytelingum's paper on the Adaptive Aggressive trading strategy \cite{aa}, which has been shown to be dominant over human traders in every scenario.

The introductory study claimed that it is dominant over the other algorithms previously mentioned but that claim has not fully held up in subsequent research by other authors. Different ratios of trading agents in the market lead to different strategies being dominant. It is not enough to pit two algorithms against each other in varying ratios and declare a winner if all pairwise scenarios are one-sided. One must consider the entire trading environment with other traders and strategies in the background, as well as different market conditions - changes in supply or demand over time or a shift in price. See \cite{aadoesntdominate} for a more detailed discussion of when and why AA does not always dominate.

The simulations discussed in this paper are based on \cite{bsepaper}, available on GitHub at \href{https://github.com/davecliff/BristolStockExchange}{https://github.com/davecliff/BristolStockExchange}.

\subsubsection{An Argument for Strategic Analysis}

There is a constant stream of research aimed at discerning market trends and the effects of various real-world phenomena on them. While this is a useful pursuit in general, there is an argument for taking a different approach to market dynamics.

An intuitive approach to the research hypothesis would be as such: imagine a rock-paper-scissors game played by thousands. If one particular player were to know that 60\% of the other players always show rock, said knowledgeable player would adapt their strategy to show more papers and such on average win more matches.

Of course, this is an extremely simplified view. Stock markets have many more variables, information and more complex strategies. These strategies often adapt to market conditions every second. Yet still, past research has shown that in certain conditions, certain algorithms are dominant over other algorithms. If one agent was given information about what algorithms the others were following - but not their internal variables - could they reliably pick a dominant algorithm for themselves? And would the addition of another agent into the trader pool lessen this dominant algorithm advantage?

The research hypothesis was that if traders were in possession of perfect strategic information, then they would be able to significantly outperform other traders not in possession of said information, in a majority of cases. In addition, this advantage was expected to decrease once noise is added to the information.

The high-level objective of this research project was to establish an upper bound for profit gained from strategic information in a market and analyse the loss in profit from noise in the information.
Specifically, the aims were:
\begin{enumerate}
\item Establish an upper bound for advantage gain-able from perfect knowledge of the strategies used by other traders in a stock market.
\item Introduce noise to the strategic information through a prediction simulation with a distorted trader strategy ratio.
\item Map the severity of the noise to the loss in advantage.
\item Examine the underlying trader ratio dynamics.
\item Check the effect of different order schedules on occurring phenomena.
\end{enumerate}

\noindent
\section{Experiments}
\noindent This section will explain the approach taken to experiment design, focusing on what factors were accurately measured and what other factors may influence the results. All parameters affecting the simulation are shown and discussed, and the code is available online on \href{https://github.com/borsim/imperfect\_oracles}{GitHub at https://github.com/borsim/imperfect\_oracles} for easy reproducibility.
As previously mentioned, all possible combinations of traders have been tried, with some constraints. Combinations included at least 1 of each trader type and had an equal number of buyers and sellers for each strategy.
These experiments are novel in the way that they focus on two previously unexplored factors in stock market dynamics: dynamic and varied supply/demand schedules; and strategic information. Neither of these factors was previously individually tested in depth and their combination brings additional interactions to be discussed as well.\\
\subsubsection{Strategic Information}
Strategic information is at the core of the research hypothesis; it was tested exhaustively. Every single possible combination of the discussed four trading algorithms was simulated for every order schedule. This ensured a good coverage of all possible scenarios of strategies in the market. Trader ratios of higher granularity that are not directly mapped out by these experiments may be approximated by interpolating from the closest ratio points. \\
\subsubsection{Supply/Demand Schedules}
Supply/demand schedules cannot be exhaustively tested: the number of possibilities is dependent on multiple potentially infinite variables. To produce a good coverage of scenarios a randomization approach was taken. Order schedules were crafted from a multi-dimensional space. They were composed of a number of sub-schedules, each with its own set of parameters. The only constraint was that supply and demand schedules do not change independently; sub-schedules on the supply and demand side were respectively equally long in duration. The number of dimensions in this space varied based on the first randomized parameter, the number of sub-schedules.

Simulations were run for a number of \textit{timesteps}. Each timestep allowed each trader to act and react once, issue orders and update internal values. The full duration was divided into intervals - this controlled the frequency of customer order replenishment and change cycles.
\begin{itemize}
    \item \textbf{Drawn once per order schedule:}
    \begin{itemize}
    \item Number of sub-schedules (integer)
    \item Duration of sub-schedules (integer)
    \item Time mode (set of possible values)
    \end{itemize}
    \item \textbf{Drawn once per sub-schedule:}
    \begin{itemize}
        \item Volatility (integer)
        \item Midprice change (integer)
        \item Step mode (set of possible values)
    \end{itemize}
\end{itemize}
The full order schedules were crafted with the sequence of steps shown in Algorithm \ref{order_schedule_algorithm}.

\subsubsection{General experiment parameters}
Simulations were ran for 240 timesteps, split into at most 8 intervals of 30 steps. These intervals each contained orders arriving to the traders. Orders prices in this interval averaged $100 \pm 40$, with each individual order being at most 60 away from the midprice. The timing of the orders is set by the deployment function. The market contained a total of 32, 16 buy-only and 16 sell-only agents. Every possible combination of traders was tried for 100 random schedules.
\begin{itemize}
    \item \textbf{Time parameters}: duration: 240, interval: 30, maximum number of sub-schedules: 8
    \item \textbf{Order schedule parameters}: midprice: 100, maximum volatility: 60, maximum midprice change: 40
    \item \textbf{Order deployment parameters}: stepmode: fixed/jittered/random, timemode: periodic/drip-poisson/drip-jittered/drip-fixed
    \item \textbf{Trader parameters}: number of buyers: 16, number of sellers: 16
    \item \textbf{Simulation parameters}: number of order schedules: 100, trials for a given trader combination and given schedule: 1
\end{itemize}
\begin{algorithm}[ht!]
Draw \textbf{\textit{timemode}} with even probability from $\{periodic, drip-poisson, drip-jittered, drip-fixed\}$ \\
Draw \textbf{\textit{$\#$ of sub-schedules}} with even probability from $\{1, 2, ..., max\_schedules\}$ \\
\For{\texttt{(num\_intervals - max\_schedules)}}{
\textbf{Extend} an evenly drawn random sub-schedule's \textbf{duration} by \textit{interval\_length}
} 
supply\_schedules = \{\} \\
demand\_schedules = \{\} \\

\For{\texttt{num\_schedules}}{
    \For{\{supply-side, demand-side\}}{
    Draw \textbf{\textit{volatility}} with even probability from $\{0, 1, ..., max\_volatility\}$ \\
    Draw \textbf{\textit{midprice\_change}} with even probability from $\{midprice - max\_change, ... , midprice + max_change\}$ \\
    Draw \textbf{\textit{stepmode}} with even probability from $\{fixed, random, jittered\}$ \\
    Set \textbf{price range lower bound} to $midprice + midprice\_change - volatility$ \\
    Set \textbf{price range upper bound} to $midprice + midprice\_change + volatility$ \\
    Set schedule \textbf{step mode} to the random \textit{step\_mode}
    Set \textbf{sub-schedule duration} to value calculated before the loop
    }
    Append \textbf{sub-schedule to} respective \textbf{list}
}

return \textbf{order\_schedule} $=$ \{\textit{timemode, supply\_schedules, demand\_schedules}\} \\
\vspace{8pt}
\caption{Creating series of random order schedules}
\label{order_schedule_algorithm}
\end{algorithm}

\subsection{Establishing the baseline}
\begin{figure}
\centering
\includegraphics[width=\linewidth]{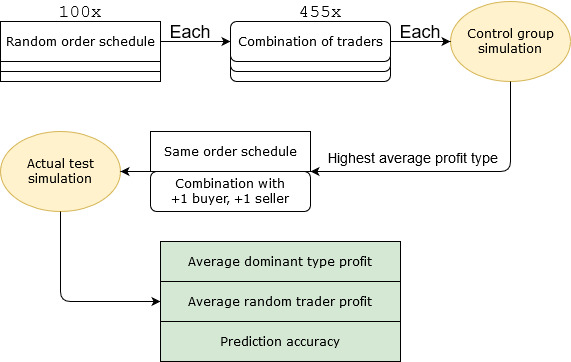}
\caption{Design of Experiment 1}
\label{fig:exp_1_design}
\end{figure}

The first experiment established a baseline, a clear limit for the maximum possible efficiency/profit achievable by a trader with access to a perfect oracle providing information about the strategies of other market participants. See Figure \ref{fig:exp_1_design} for an overview of the experiment design.

A ``control group'' market simulation was performed for a given order schedule and a given combination of traders. The simulation returns the average profit achieved by traders of particular types. This serves as the oracle. The trader type with the highest average profit was deemed ``dominant''. There was no distinction between buyers and sellers for the purposes of strategy dominance, their account balances were pooled to obtain the average. An additional trader of this dominant type was added to the buyer and the seller pool - this was to take into account the effect of an intelligent agent on the market. Simulating a market then counting the best outcome will be, simply put, the best. Simulating a market and slightly changing the conditions this way will show if the observations have actionable value and whether something optimal in one setting could remain close to optimal in another. \\ The second simulation was performed with the thusly expanded pool, this time to obtain the real data.

Between the two simulations trader strategy ratios and overall supply/demand patterns were shared. Particular orders may be different if the stepmode contained randomness - i.e. it was not fixed. The sequence of order arrivals may also differ. \\
The expected result of the experiment was that traders following the optimal strategy determined in the control simulation would usually dominate in the repeated simulation. The concrete amount of profit earned of course depended on the supply/demand schedule the simulation was performed with. For this reason the target end result of the experiment was a ratio; a multiplier on the average trader's profit.

\subsubsection{Results of Experiment 1}
\begin{figure}[ht!]
    \centering
    \includegraphics[width=0.9\linewidth]{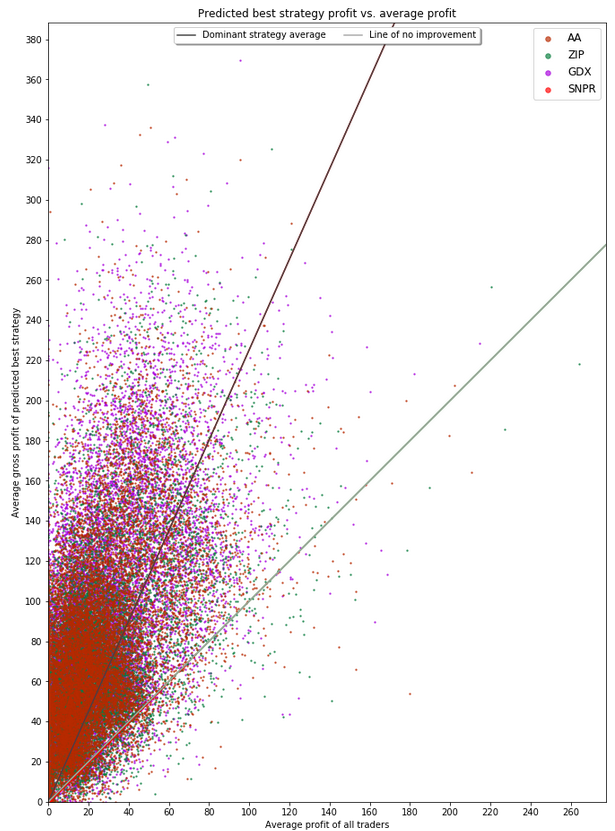}
    \caption{Overall results of Experiment 1}
    \label{fig:experiment1base}
\end{figure}
The results of Experiment 1 can be viewed in this representation on Figure \ref{fig:experiment1base}. It shows a comparison of how the strategy predicted as dominant fared in the second simulation with the expanded trader pool. The axes mark the gross average profit per trader of the predicted dominant type and the colour of points indicates which strategy it is. For added visibility the ``breakeven line'' is also plotted - on this line the predicted best type trader earned exactly as much as the market average.

The data confirmed the abovementioned expectations. A number of points fell below the breakeven line but the majority are above and the data could be reasonably fit on a line. Predicting the best strategy would, indeed, on average keep a trader above market average.

\subsection{Experiment 2}
Experiment 2 attempted to measure the effect of the noise described in the experiment design section. A similar overall design was adopted for Experiment 2. The significant difference was the presence of a distortion between the prediction and the actual test ratios. Additionally, the number of order schedules and trader combinations varied; this was only a numerical difference and does not have an effect on methodology. 

Experiment 2 was not immediately conclusive. Smaller scale trials were run first to confirm the presence (or lack thereof) of correlation before committing a large amount of computational resources. Methodology was then adapted to reduce the noise in the resulting data. Finally a large trial was performed to provide more accurate numerical output showcasing the merit of the previous changes. All components of Experiment 2 had the market session duration extended from 240 to 330 timesteps.

\subsubsection{Definition of noise used in this experiment}
One could have selected a different definition of noise and it would not influence the validity of the results derived here. It is possible to conduct analysis on different noise patterns by looking purely at the data points and connecting any two as a prediction and result into a different noise definition. The particular noise function for this research was chosen for simplicity and intuitive ease. It is parametrized by a single parameter $p$. Each trader in the ``real'' point has $p$ probability of being viewed as one following a strategy that is not the same as their true strategy. The ``mistaken'' strategy is chosen with an even probability from the strategies present in the market, excluding the ``true'' strategy. As such $p$ scales the expected distance of the prediction point from the real point and the direction was chosen uniformly from the set of available points. By this definition maximum noise was achieved when all traders have an equal probability of being seen as any strategy. $$p_{max} = 1 - (\frac{1}{|S|})$$ where $S$ is the set of possible strategies and $|S|$ denotes the number of elements in $S$.

\subsubsection{Inconclusive Trial}
The first trial closely followed Experiment 1. It also involved $100$ random order schedules. Each order schedule was assigned a noise $p$ with the first having $0\%$, the second $0.75\%$ following to the 100th having $(75-0.75)\%$ and each of these schedules had all trader combinations trialed once. The line fit coefficient was less than $0.0003$ away from horizontal and visual observation did not reveal strong trends that were covered with noise or outliers. This means that there was no evidence that in this case the prediction conferred a valuable piece of information.

The number of incorrect predictions - cases where the predicted best type did not earn the highest average profit - was expected to have an upwards slope as noise in the prediction ratios increased. While it did have an upwards slope, the data surrounding it was very noisy and the change itself was fairly small compared to the full set. With a coefficient of $0.01$, at the highest noise level (where on average each trader is assigned a type through a roll of a $4$ sided die) on average $7.5$ more incorrect predictions were made. This was a change of $1.5\%$ when in the context of a full set of 455 prediction-result pairs, the change being a result of going from zero noise to the maximum possible noise. Combined with the previous result on profits this suggested that these predictions are just not particularly good. Over half of them were incorrect from the outset.

\subsubsection{Reduced Randomness in Order Schedules}
A followup mini-experiment aimed to check whether the randomisation of order schedules had too much of an effect on the outcomes of the previous, inconclusive trial. For this experiment the order schedule was set to a very simple base case reminiscent of those of Vernon Smith. The supply and demand ranges had equal limits, prices were allocated at even steps in these ranges and resupplied to all traders periodically. Save for the order-trader allocations and the order of incoming orders from the set everything else was deterministic. This schedule was similarly sampled at $10$ distinct noise probabilities.

The line's slope coefficient was $-0.001$ (per $1\%$ of probability). While this was an order of magnitude greater than that of the previous experiment it was still lacking the expected, slightly more marked decrease. While the overall number of incorrect predictions was lower in the very simple and deterministic schedule, the environment was still noisy and the smaller number of points in this mini-experiment even produced a downward trend. Considering how poor the fit was it should just be regarded as no evidence for a correlation.

\subsubsection{Averaged Samples}
The next mini-experiment focused on the low prediction accuracy shown so far. It sampled the simple order schedule at $10$ evenly spaced probabilities and introduced a large change in how predictions were made. More than one prediction was made each time - in this case, $10$. The predicted best trader was the one that has the highest average profit in the sum of the $10$ predictions combined. The real data trials were done in a similar fashion; the final value for a prediction-real pair was the average profit in the $10$ trials. Note that each prediction used the same (noisy) trader ratio combination. Trader ratios were not re-randomized in-between predictions.

As a result of these changes the range of profits was much narrower. Because multiple trials were averaged with a simple mean calculation that is linear in its treatment of outliers, the averaged values were lower as well when contrasted to the least-squares fit of the previous dataset. The data was far more compact with a range of just $0.5-3$ instead of $0-10$, indicating significantly improved prediction accuracy. Instead of profit multiplier values of $5$ and beyond, no single averaged-trial went above $2.5$. However, the slope of the profit line was still very close to horizontal with a coefficient of $0.001$. Despite that, this coefficient was objectively a more useful value due to how the narrower and more regular data range led to a stronger implication of a cause and effect rather than random noise.

Prediction accuracy further supports the above argument. The previous prediction accuracy line fit is of very poor quality and as such exact metrics like the residuals of the fit were meaningless in context. The line fit of this experiment, however, was far better. It shows the expected clear upward trend - more noise in the prediction resulted in more wrong predictions. This trend was also significantly higher in impact than the previous existing trends. Going from $277$ wrong predictions at no noise to $297$ at maximum noise it nearly tripled the previous prediction error change with a better fit. \\
For the large scale simulation repeated predictions and real trials appeared to be a must.

\subsubsection{Large-scale Experiment: repeated predictions \& randomized schedules}
One additional change was made in addition to the previously discussed ones. Due to its low performance and unlikeliness of being the best on its own merit, SNPR traders were taken out of the pool of possibilities. This resulted in a number of changes to the starting parameters of the simulation: 
\begin{itemize}
    \item The total number of participating traders decreased with the number of available strategies to $4*3 = 12$ for the prediction trials and $12+2$ for the real trials
    \item The total number of possible trader combinations dropped from $455$ to $55$
    \item The maximum possible noise probability lowered to $66.6\%$
\end{itemize}

Due to how passive and nonadaptive SNPR traders have proven to be, this change had an additional beneficial effect. The three advanced algorithms were now in closer contest with fewer bystanders, meaning that the profit ratio being measured is closer to 1. Previously all of them had a fair amount of extra profit just by taking advantage of bad SNPR trades. 

This large trial was done on $10$ different and randomized order schedules. In one of the schedules the randomizer produced an unfavourable schedule, resulting in few to none trades in most of the trials and as such the data from that was discarded. The other schedules were tested at $14$ distinct noise probabilities in the range of $0\%-65\%$ with increments of $5\%$. Each probability point was tested for all 55 trader combinations and each trader combination had a set of $50$ prediction subtrials and $50$ real subtrials.

Before the summaries relating to the main hypothesis of this research are presented, a slight correction in evaluation methods must be made. Previous, less exhaustive experiments did not show the phenomena presented below in an impactful way when plotting profit so the credibility of using least squares line fits for them has not changed. For this subsection however, the least squares fit on its own was no longer an accurate enough tool. The market average profit should, under every condition, stay constant. In cases where a line fit would indicate this not being the case, the line fit is wrong. The market average does not change with noise and it was taken from all trader combinations equally for each probability. The target of analysis is the difference in slopes between the two lines of market average and enhanced trader average. Analysis of cases where the market average was not flat was problematic - though less so than it appears due to the comparison being point-wise rather than line-wise. Still, the least-squares fit in some complex schedules should be taken with a grain of salt.

The observed phenomena still hinted at being the closest to fitting on a line. Fits of higher degree polynomials were attempted but the higher-rank coefficients were close to $0$. Even after increasing prediction accuracy some outlier removal steps still had to be taken, especially with the re-inclusion of more complex order schedules. The data range has a complex structure with traders sometimes earning extraordinarily high profits - outliers on the upper end were far more common than those on the lower end. Outliers were removed with a method common in statistics but with slightly different parameters. Generally, points outside $1.5$ times the inter-quartile range are considered outliers. For the purposes of not losing too much important data - it should be noted that even these outliers awerere a result of $50 + 50$ individual data points - this interquartile range requirement was loosened. The central range was the interdecile range (between $10\%$ and $90\%$). The multiplier for acceptable points was $1$ times this range from the edges.

\begin{figure}
    \centering
    \includegraphics[width=\linewidth]{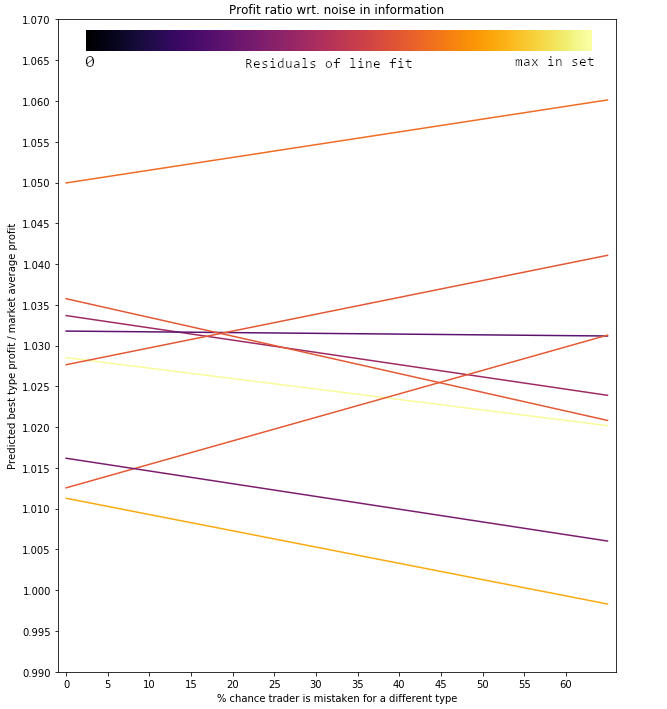}
    \caption{Profit trends for 9 order schedules}
    \label{fig:experiment2bigalllines}
\end{figure}

\begin{figure}
    \centering
    \includegraphics[width=\linewidth]{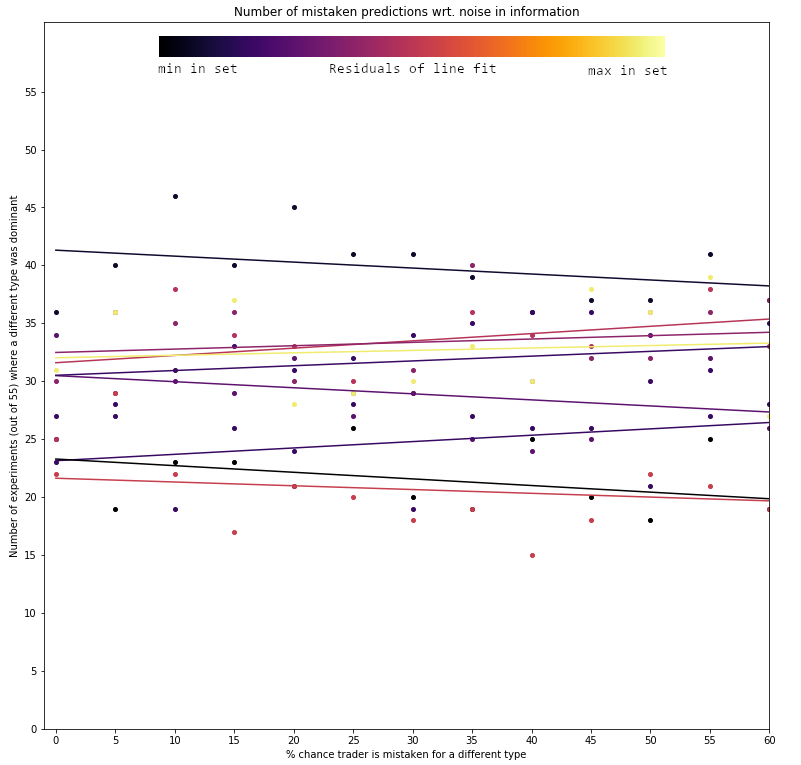}
    \caption{Prediction accuracy for 9 order schedules}
    \label{fig:experiment2bigpredictions}
\end{figure}

Figure \ref{fig:experiment2bigalllines} shows the line fits on all schedules. The colour of the line indicates how good the line fit is; a darker colour has closer to 0 residuals. The lightest colour was set as the maximum residuals out of the 9 line fits.
The majority of trends displayed a clear downwards slope, among which were the three closest fits. One third of the studied order schedules displayed an upwards slope of moderate uncertainty. These trends are closely paired with the prediction accuracy graphs in Figure \ref{fig:experiment2bigpredictions}, which shows four lines with a downwards slope where upwards was the expected direction - more noise meaning more prediction errors. Three of those coincided with the upwards profit schedules and one poor line fit of prediction accuracy had a downward slope on the profit graph.

\begin{figure}
    \centering
    \includegraphics[width=0.65\linewidth]{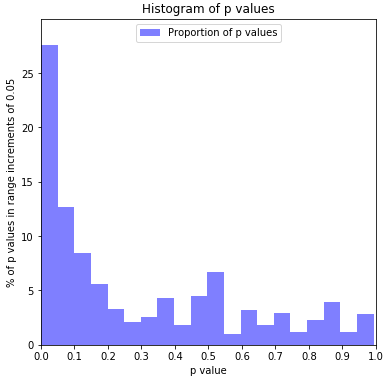}
    \caption{P values of Wilcoxon tests}
    \label{fig:experiment2bigpvalues}
\end{figure}

The overall result of the large-scale experiment supported the research hypothesis. The majority of order schedules tested show an above-average profit earned from good predictions that steadily decreased over increasing prediction noise in strategic information. Figure \ref{fig:experiment2bigpvalues} is a visualization of distribution differences in the trial profits. It presents comparisons through nonparametric Wilcoxon tests of the distributions involved in the nine order schedules.

\begin{figure}
    \centering
    \includegraphics[width=0.65\linewidth]{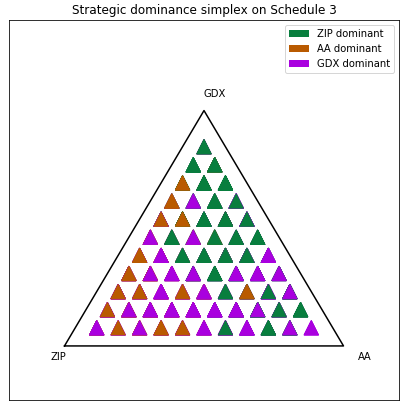}
    \caption{Well-predictable order schedule}
    \label{fig:experiment2bigsimplexes1}
\end{figure}

\begin{figure}
    \centering
    \includegraphics[width=0.65\linewidth]{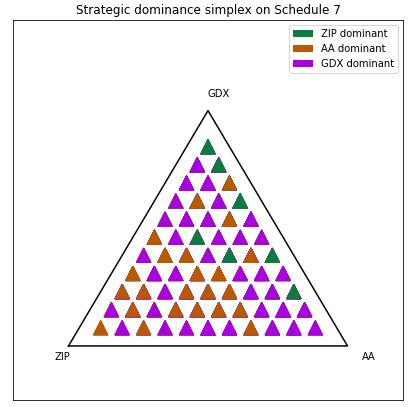}
    \caption{Badly-predictable order schedule}
    \label{fig:experiment2bigsimplexes2}
\end{figure}

Individual data points are a pairwise comparison between the noiseless prediction-result distribution and a noisy prediction-result distribution. Small p values indicate strong certainty that the samples were drawn from different distributions. A large portion of these tests indicate that the samples with noise involved can usually be distinguished from samples without prediction noise.
A possible explanation for the uncertain predictions and upwards profit trends could be that certain order schedules spawned more complicated strategic dynamics. For the majority of schedules the addition of two extra traders of a kind did not nullify the advantage of having had access to predictions. However, for the schedules that show negative correlations, this alteration in trader ratios might have ended up working against the traders with access to an oracle, pushing them over the boundary where the predicted strategy was no longer the optimal one. Such a scenario could have arisen when most of the strategic landscape was dominated by a single trader with very small pockets of other types. Figure \ref{fig:experiment2bigsimplexes1} and Figure \ref{fig:experiment2bigsimplexes2} show a comparison between Schedule 3 - where predictions performed well - and Schedule 7 where they did not.

\section{Conclusions}
\label{sec:conclusion}

\noindent  The fairness and regulatory tools of financial markets draw notable attention, especially after recessions like the 2008 housing crash or the fallout of the 2020 COVID pandemic. It is vital to have scientifically estimated bounds for what advantage is reasonable as a function of access to information. This exploratory study suggests an approximate $10\%$ profit advantage difference between perfect and no information in most cases. Excessive advantage could prove to be a strong hint towards requiring further, manual investigation of a market participant.

As well as confirming the initial hypothesis, this research explored the most significant factors influencing market predictions in a simulated environment. Future research now has a proven process to use as a basis, with knowledge of the primary obstacles in assessing predictions, such as the need for an average prediction set, as opposed to perfect knowledge of the current market state. Some failure cases of the proposed prediction system were also unveiled, where a counter-intuitive correlation was present. The failure cases were supported in evidence of their differing nature from two independent sources of experiment data, profit ratio and prediction accuracy. Following work can either avoid said failure cases or delve deeper into why they produce the results in question.

This project extended the Bristol Stock Exchange simulator with more functionality. These extensions are open for anyone to copy, use or modify. The data pipeline of the analysis is also published online and may be used directly on any experiment data file with no additional steps.

All data points in the data files may be reinterpreted in their relation to each other. The data obtained remains useful as one may perform an analysis of a different noise phenomenon without the need to re-generate hundreds of thousands of lines of data.

\noindent
\bibliographystyle{apalike}
{\small
\bibliography{ImperfectOracles}}
\pagebreak
\appendix
\section{Appendix: Additional Figures}
\begin{figure}[h!]
    \centering
    \includegraphics[width=0.64\linewidth]{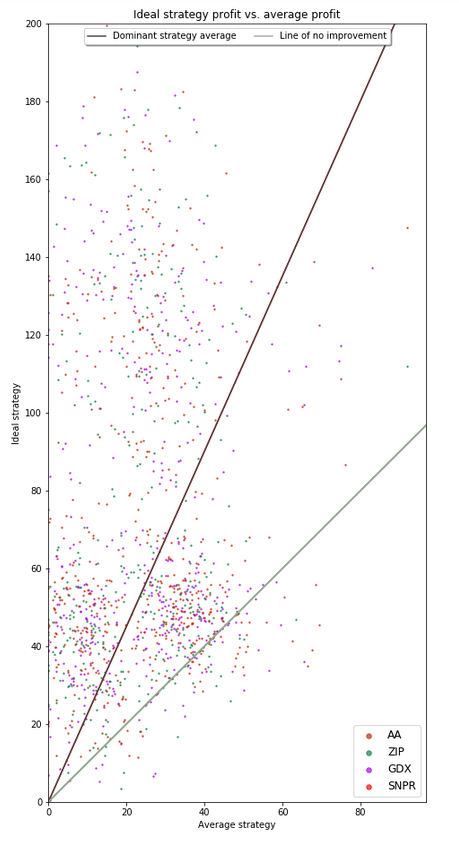}
    \caption{Three order schedules in Experiment 1}
    \label{fig:experiment1fewschedules}
\end{figure}
\begin{figure}
    \centering
    \includegraphics[width=0.95\linewidth]{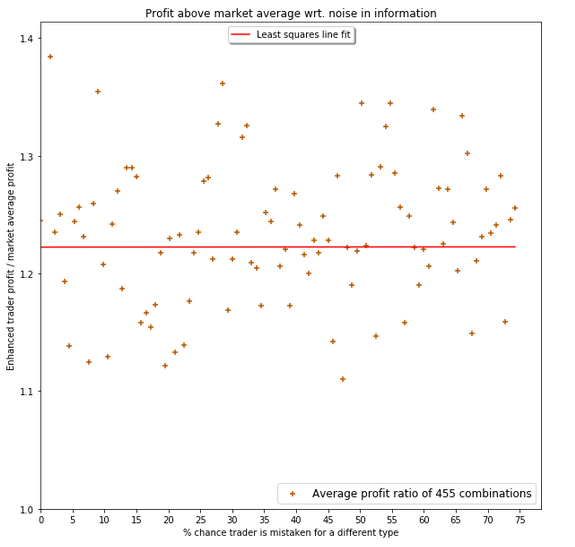}
    \caption{Profits in the initial Experiment 2}
    \label{fig:experiment2initial}
\end{figure}
\begin{figure}
    \centering
    \includegraphics[width=0.95\linewidth]{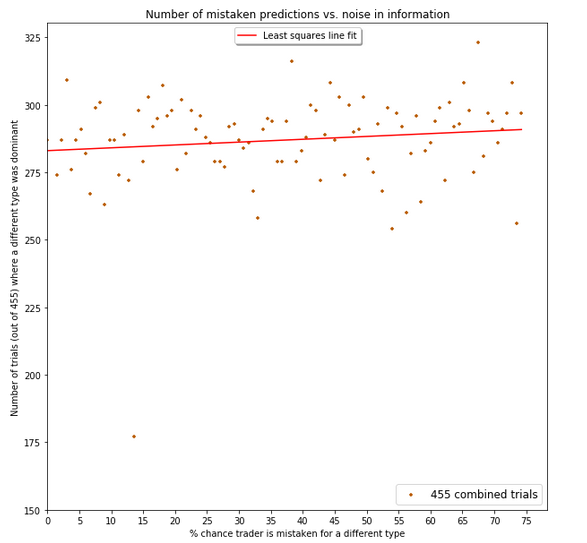}
    \caption{Prediction accuracy in the initial Experiment 2}
    \label{fig:experiment2initialpredictions}
\end{figure}
\begin{figure}
    \centering
    \includegraphics[width=0.95\linewidth]{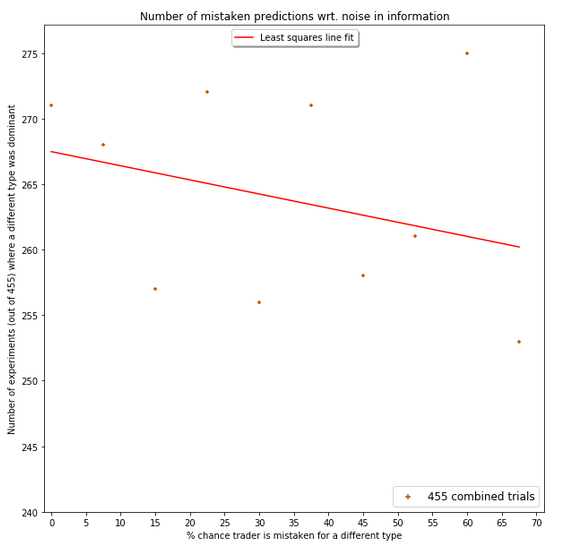}
    \caption{Prediction accuracy on the simple order schedule}
    \label{fig:experiment2simplepredictions}
\end{figure}
\begin{figure}
    \centering
    \includegraphics[width=0.95\linewidth]{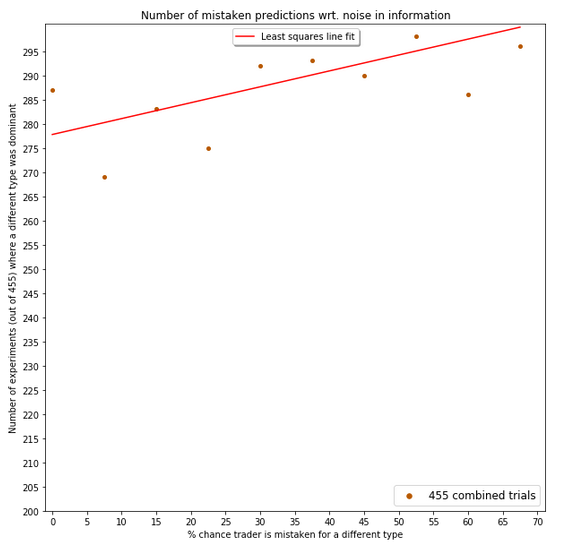}
    \caption{Prediction accuracy on simple order schedule \& multiple subtrials}
    \label{fig:experiment2multitrialpredictions}
\end{figure}
\begin{figure}
    \centering
    \includegraphics[width=0.95\linewidth]{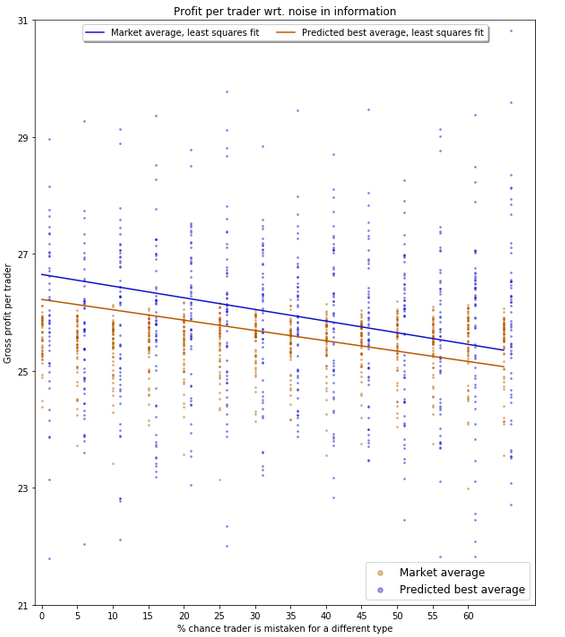}
    \caption{Problems with least squares fitting}
    \label{fig:experiment2biglinefit}
\end{figure}
\end{document}